# Cycles of paleomagnetic activity in the Phanerozoic

A. Yu. Kurazhkovskii[a], N. A. Kurazhkovskaya[a], B.I. Klain[a]

[a]*Geophysical Observatory Borok, Schmidt Institute of Physics of the Earth of the Russian Academy of Sciences, Borok, Yaroslavl Region, 152742, Russian Federation*

**Abstract**

Quasi-periodic changes of the paleointensity and geomagnetic polarity in the intervals of 170 Ma to the present time and of 550 Ma to the present time were studied, respectively. It is revealed that the spectrum of the basic variations in the paleointensity and of the duration of the polar intervals is discrete and includes quasi-periodic oscillations with characteristic times of 15 Ma, 8 Ma, 5 Ma, and 3 Ma. The characteristic time of these quasi-periodic changes of the geomagnetic field at the beginning and at the end of the Phanerozoic differed by no more than 10%. The spectral density of quasi-periodic variations of the geomagnetic field changed cyclically over geological time. The relation between the behaviors of the amplitude of paleointensity variations, the duration of the polar intervals, and their spectral density was shown. Quasi-periodic variations of the paleointensity (geomagnetic activity) had a relatively high spectral density in the interval of (150 - 40) Ma (in the Cretaceous - Early Paleogene). In this interval, both the amplitude of paleointensity variations and the duration of polar intervals increased. In the intervals of (170 - 150) Ma and of 30 Ma to the present, a quasi-periodic variation in the paleointensity practically did not detect against the background of its noise variations. At the same time, the amplitude of the paleointensity variations and duration of polar intervals decreased. An alternation of time intervals in which the paleointensity variations acquired either a quasi-periodic or noise character took place during the geomagnetic history.

## 1. Introduction

Conclusions about the processes occurring in the Earth's core and the mechanism of their connection with cyclic lithospheric processes are based mainly on data on the features of changes in the ancient geomagnetic field, for instance (Khramov et al., 1982; Larson and Olson, 1991; Hounslow et al., 2018). Paleomagnetic data indicate that the paleointensity (PINT database (2015.05); Heller et al., 2003) and the duration of geomagnetic polarity intervals, for example (Gradstein et al., 2008), changed cyclically during geomagnetic history. At the same time, information on the spectrum of characteristic times and amplitude of changes in parameters that characterize the geomagnetic field is still fragmentary and not systematized. Thus, cyclical changes of the frequency of geomagnetic reversals (Khramov et al., 1982) and the paleointensity (Bol'shakov and Solodovnikov, 1981) with large characteristic times (about 200 Ma) are well known. Changes of the paleointensity and frequency of geomagnetic reversals in these geomagnetic cycles are inversely related (Kurazhkovskii et al., 2010). A similar conclusion on the relationship between the behavior of the paleointensity and the frequency of geomagnetic reversals was made later by Kulakov et al. (2019).

The existence of cyclic changes of frequency of the geomagnetic reversals in the Cenozoic with characteristic times of the order of 15 Ma was showed by Mazaud et al. (1983). However, cyclic changes inherent in both the behavior of paleointensity and the frequency of geomagnetic reversals with characteristic times of less than 100 Ma remain practically has not been studied.

The results of determining of the geomagnetic field intensity from sedimentary and thermomagnetized rocks obtained in recent years made it possible to continue studies of the patterns of the paleointensity changes. In this study a search of common quasi-periodicities inherent of changes of the intensity and polarity of the geomagnetic field was carried out based on the analysis of various arrays of paleomagnetic data.

## 2. The analyzed data and method

In this study, we used the results of paleointensity definitions obtained from marine and oceanic sediments, as well as from thermomagnetized rocks. The results of paleointensity determinations by sedimentary rocks were taken from (Oxhneiser et al., 2013; Yamamoto et al., 2014; Yamazaki and Yamamoto, 2018; Kurazhkovskii et al., 2020). The paleointensity definitions by thermomagnetized rocks were taken from the PINT database (2015.05) (http://earth.liv.ac.uk/pint/), which was described by Biggin et al. (2010).

The results of paleointensity determinations by sedimentary rocks allow for analysis its behavior in the intervals of (8 - 0) Ma, (19.5 - 12) Ma, (42 - 23) Ma and (170 - 23) Ma. The sediments which were used to determine the paleointensity in the intervals of 42 Ma to the present time (Ohneiser et al., 2013; Yamamoto et al., 2014; Yamazaki and Yamamoto, 2018) and (167 - 23) Ma (Kurazhkovskii et al., 2020) were sampled in different regions. This allowed for comparison the results of paleointensity determinations in the interval of (42 - 23) Ma, which were obtained by deposits of different regions. Paleointensity data on sedimentary rocks were calibrated using the PINT database (2015.05). The calibration procedure for paleointensity data is described in sufficient detail by Kurazhkovskii et al. (2011).

For analyzing of changes of the duration of the polar intervals, we used the Geologic Time Scale 2008 (Gradstein et al., 2008), as well as the general scale of geomagnetic polarity (Zhamoida et al, 2000) with adjustments in the Jurassic and Cretaceous intervals which given in works (Guzhikov et al., 2007; Pimenov and Yampolskaya, 2008). These geomagnetic polarity scales were used to assess the effect of paleomagnetic data detail on the results of studying of cyclical changes of the geomagnetic polarity. The behavior of geomagnetic polarity was analyzed on the basis of the data on the duration of polar intervals. This approach does not require data averaging, and it accurately reflects the temporal distribution of geomagnetic reversals which corresponds to the used the scales of geomagnetic polarity.

According to the scales (Zhamoida et al, 2000) and (Gradstein et al., 2008), the average duration of the polar intervals in the Phanerozoic was about 400 ka. This allowed for searching cyclical changes of the duration of polar intervals with a resolution of several million years. In this study, we tried to determine the spectrum of cyclical changes of the duration of polar intervals and paleointensity with a resolution of up to 2 Ma.

The spectrum of quasi-periodic changes of the paleointensity was studied using wavelet analysis (Astaf'eva, 1996). For this purpose, the paleointensity curve obtained from sedimentary rocks was scanned and digitized with resolution of 50 ka, which is sufficient for study variations with characteristic times starting from 500 ka upwards. Effective algorithms for wavelet analysis were realized in the software (MATLAB, 2018). The width of the window for the spectral density estimation is proportional to the temporal length of the wavelet. The wavelet characteristics are defined by the minimum time resolution (digitization step) and the length of the studied series. The optimal values of the wavelet parameters were computed in the software (MATLAB, 2018) automatically and used as default.

In addition, in this study for search of hidden quasi-periodicities in the behavior of paleointensity and the duration of polar intervals, we used Fourier analysis, the method of which is described by Serebrennikov and Pervozvansky (1965). For Fourier analysis of data (paleointensity - H, or durations of polar intervals - Δt), the values of which are not specified at equidistant points of the investigated time interval, we approximate the sequence by a series:

$$F(t) = \sum_{j=1}^{N} s_j \delta(t - t_j). \qquad (1)$$

Here $F$ is the analyzed series of data characterizing the ancient geomagnetic field, $t_j$ is the time to which the number $s_j$ is assigned, and $N$ is the total number of such intervals. We expand the function $F(t)$ into the Fourier integral:

$$F(t) = \int_{-\infty}^{\infty} F_\omega \exp(-i\omega t) \frac{d\omega}{2\pi}. \qquad (2)$$

From the last expression, taking into account (1), we find the spectral component $H_\omega$

$$F_\omega = \sum_{j=1}^{N} s_j \exp(i\omega t_j).  \qquad (3)$$

We used formula (3) for obtaining of the spectrum of the studied below series.

## 3. Cyclic changes of the paleointensity

The results of determining of the geomagnetic field intensity in the interval of the Middle Jurassic – present times by sedimentary and thermomagnetized rocks is shown in Fig. 1a and Fig. 1b.

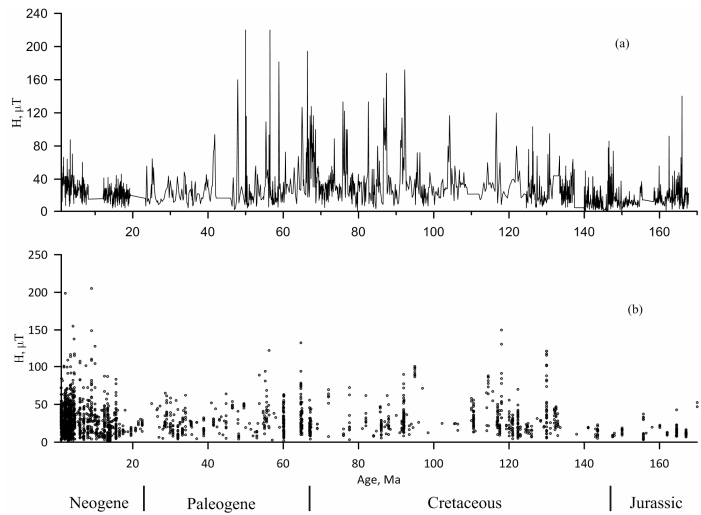

Fig. 1. Behavior of paleointensity in the last 170 Ma: (a) by sedimentary (Oxhneiser et al., 2013; Yamazaki and Yamamoto, 2018; Kurazhkovskii et al., 2020) and (b) thermomagnetised rocks from the PINT database (2015.05).

As can be seen from the Fig. 1, short-term increases of the paleointensity (bursts) were by the most significant geomagnetic events associated with the energy of the geomagnetic field. During the bursts, the paleointensity increased several times compared to its average value (29 µT in the interval of (170 - 23) Ma). The behavior of paleointensity is characterized by an alternation of short series of paleointensity bursts with intervals of a relatively quiet geomagnetic field. We consider such changes in the geomagnetic field as cycles of the paleointensity or geomagnetic activity. In these cycles, the amplitude of the paleointensity bursts changed with geological time. The maximum amplitude of the paleointensity bursts were observed in the Cretaceous - Early Paleogene. After boundary of 40 Ma (Fig. 1), the amplitude of paleointensity bursts decreased.

Figure 2 shows the wavelet diagrams of paleointensity variations in the interval of 170 Ma to the present time.

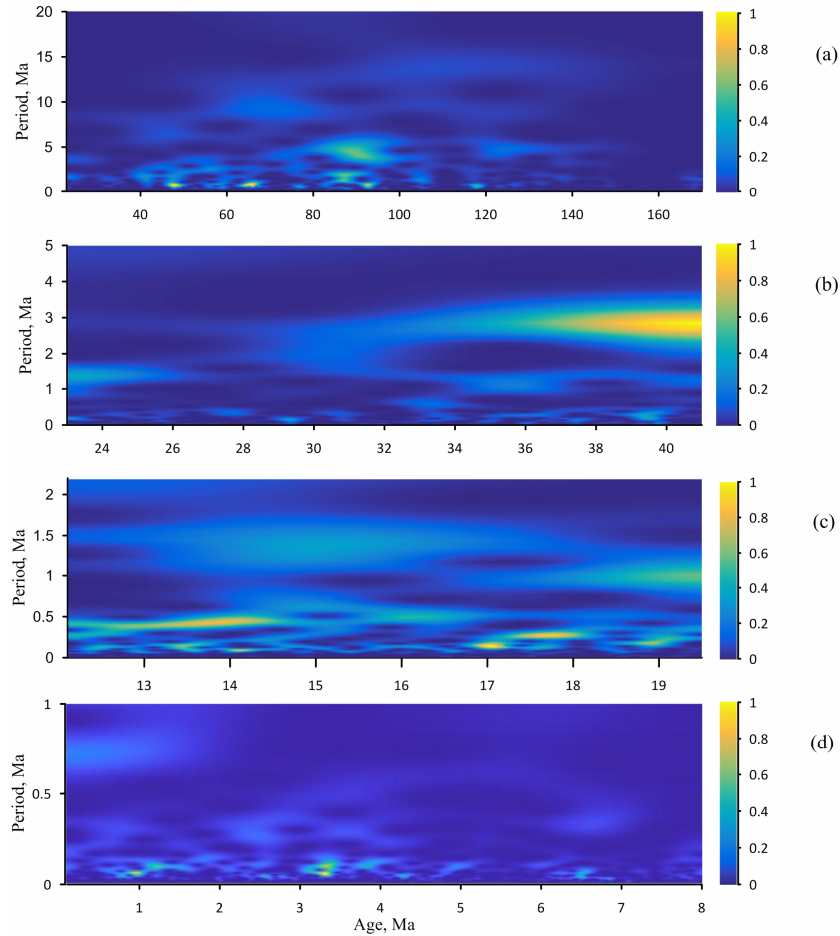

Fig. 2. Wavelet diagram of changes of paleointensity obtained by analysis by sedimentary rocks in the intervals: (a) (167 - 23) Ma (Kurazhkovskii et al., 2020), (b) (42 - 23) Ma (Yamamoto et al., 2014), (c) (19.5 - 12) Ma (Oxhneiser et al., 2013), (d) of 8 Ma to the present (Yamazaki and Yamamoto, 2018).

As can be seen from the results of the wavelet analysis (Fig. 2a), the spectral composition of paleointensity variations changed with geological time. For instance, in the Jurassic (170 - 150) Ma, no obvious periodicities are found in the spectrum of paleointensity variations. In the Cretaceous - the beginning of the Paleogene, the spectrum of paleointensity changes becomes discrete. Quasi-periodicities with characteristic times of the order of 15 Ma, (8 - 10) Ma, 5 Ma, 3 Ma, and 1 Ma are appeared in this interval. In the Paleogene, the spectrum width of quasi-periodic variations in paleointensity decreased. In the interval of (42 - 23) Ma, the characteristic times of quasi-periodic changes of paleointensity were 3 Ma and 1 Ma (Figs. 2a, 2b). At the same time, paleointensity variations with characteristic times of more than 2 Ma deceased after boundary of 30 Ma (Figs. 2b, 2c, 2d).

According to the results of Fourier analysis (Fig. 3a), in the interval of (170 - 23) Ma the spectral density maxima corresponded to changes of paleointensity with characteristic times of 5 Ma and 8 Ma. Similar maxima in the spectrum of paleointensity changes were obtained for thermomagnetized rocks from PINT (2015.05) (Fig. 3b).

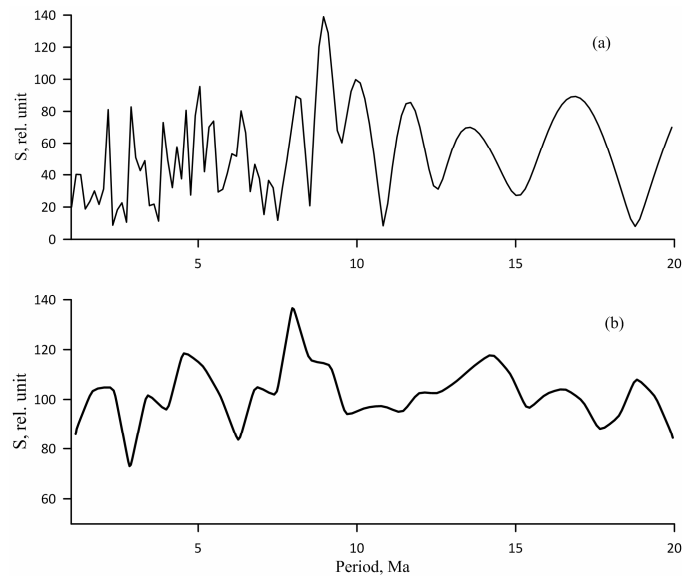

Fig. 3. Spectrum of quasi-periodic changes of paleointensity obtained as a result of Fourier analysis of data (a) on sedimentary (Kurazhkovskii et al., 2020) and (b) thermomagnetized rocks from PINT (2015.05).

As can be seen from the results of the wavelet analysis (Fig. 2), the time intervals in which paleointensity variations were either quasi-periodic or noisy were alternated. In the Cretaceous and early Paleogene the spectrum of paleointensity variations was discrete (Fig. 2a). For determining the spectrum of paleointensity changes in more distant intervals of geological time, there is still little paleomagnetic data. Nevertheless, by Kurazhkovskii et al. (2020) it was shown that paleointensity bursts and associated with it cyclical changes with characteristic times of several million years took place in the Permian - early Triassic. This provide a basis for the assumption that in the Permian (270 - 250) Ma the spectrum of paleointensity changes was also discrete. Chaotic state of paleointensity changes are noted in the Jurassic and Neogene. Thus, over the course of geomagnetic history, time intervals in which changes of the paleointensity were either quasi-periodic or chaotic alternated. The available materials provide a basis for the conclusion that the intermittency of intervals with quasi-periodic and chaotic changes in paleointensity is a cyclic process. According to our estimates, its characteristic times are close to the durations of the known cycles of changes in the frequency of geomagnetic reversals in the Phanerozoic (about 200 Ma).

We have revealed two types of cyclical changes in paleointensity: cyclical changes in its value and cycles during which the structure of its variations changed. The spectrum of paleointensity variations contains quasi-periodicities with characteristic times of about 15 Ma, 8 Ma, 5 Ma, and 3 Ma. The paleointensity variations reflect changes in the energy of the geomagnetic field. In this regard, we consider cycles of paleointensity changes as cycles of geomagnetic activity. The most significant changes in geomagnetic activity occurred in cycles with characteristic times of the order of 5 Ma and 8 Ma (Fig. 2a, Fig. 3a, 3b). Increases or decreases of the duration of geomagnetic cycles were accompanied by decreases in their spectral density. The estimates of the cycles of geomagnetic activity made from thermomagnetized (Fig. 3b) and sedimentary rocks (Fig. 3a) coincided with an error of the order of 1 Ma.

The most significant changes in the structure of paleointensity variations occurred near the boundaries of 150 Ma and 30 Ma. Until 150 Ma, no quasi-periodicity was found in the spectrum of paleointensity changes. In the interval of (150 - 30) Ma, at the behavior of paleointensity revealed a series of quasi-periodicities with characteristic times from 15 Ma to 3 Ma. After 30 Ma, quasi periodic changes of paleointensity with characteristic times of more than 2 Ma deceased.

**4. Changes of duration of the polar intervals**

The dynamics of changes in the duration of the polar intervals at the end of the Phanerozoic (last 170 Ma) and over the entire Phanerozoic (of 550 Ma to the present) shown in Figs. 4a - 4c. Depending on the details of the used data, the distributions of the polar interval durations in the last 170 Ma differ somewhat (Figs. 4a, 4b). Also noteworthy is the different detail (Fig. 4c) of the study of the geomagnetic polarity regime at the beginning (550 - 170) Ma and at the end of the Phanerozoic (of 170 Ma to the present). The division of the Phanerozoic into two unequal parts was determined by differences in the details of paleomagnetic data in these intervals of geological time.

According to all the analyzed data, the most significant events in the behavior of geomagnetic polarity were long polar intervals with characteristic times of more than 1 Ma. In the interval of (150 - 30) Ma, the cyclicity is visually detected in the behavior of the polarity regime (Figs. 4a, 4b). This cyclicity is determined by the intermittency of long geomagnetic intervals by a series of short intervals of geomagnetic polarity.

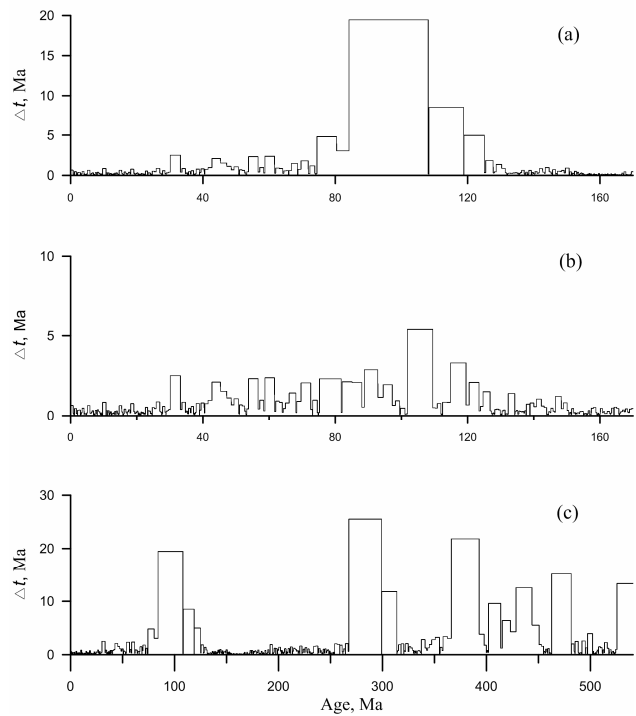

Fig. 4. Changes of the durations of polar intervals: in the last 170 Ma (a) (Gradstein et al., 2008) and (b) (Zhamoida et al, 2000; Guzhikov et al., 2007); (c) in the interval of 550 Ma to the present (Gradstein et al., 2008).

Regardless of the details of the used data, the most obvious alternation of short and long (more than 1 Ma) polar intervals is found in the interval of (125 - 30) Ma (Figs. 4a, 4b). At the same time, up to boundary of 115 Ma, the duration of the polar intervals increased. After 30 Ma to the present, the durations of the polar intervals decreased. The longest intervals of geomagnetic polarity were between 125 Ma and 75 Ma.

During frequent reversals of the geomagnetic field (170 - 150) Ma, no cyclicity is found in the behavior of the geomagnetic polarity. This can be in many respects due to the presence of zones of anomalous polarity (Pimenov and Yampolskaya, 2008), in which it is difficult to determine both the duration of individual polar intervals and their number. Thus, cyclical changes in the duration of polar intervals were visually revealed after boundary of 150 Ma, as well as earlier than of 170 Ma (Fig. 4c) and were not detected between 170 Ma and 150 Ma (at the end of the Middle - the beginning Late Jurassic). In the geomagnetic history, there was an alternation of time intervals, in which the cyclical behavior of the geomagnetic polarity regime was either found or not.

Fourier analysis of the scale (Gradstein et al., 2008) was showed (Figs. 5a, 5b) that the behavior of the polar interval durations is characterized by quasi-periodic changes with characteristic times of about 3 Ma, 5 Ma, 8 Ma, and 15 Ma.

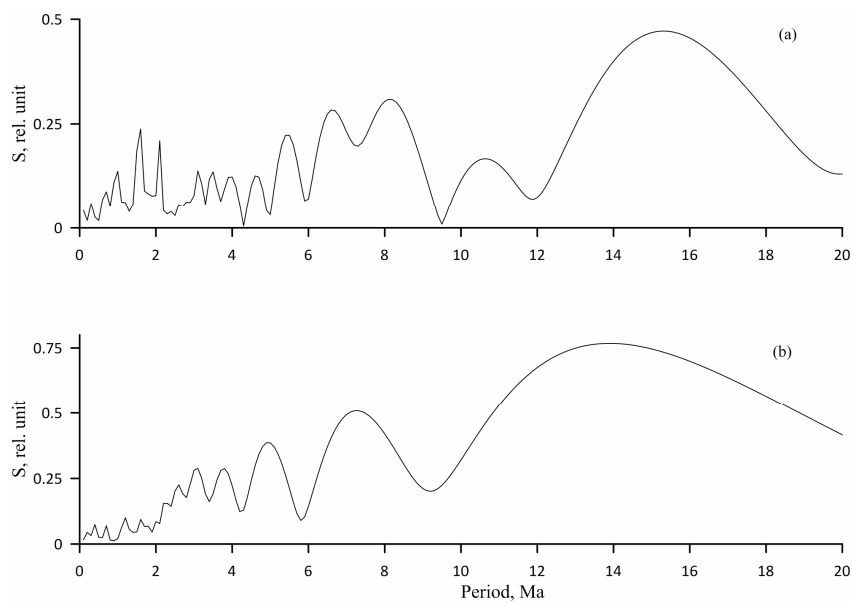

Fig. 5. Spectrum of the duration of polar intervals: (a) at the beginning (550 - 170) Ma and (b) at the end (of 170 Ma to the present) of the Phanerozoic.

At the beginning (550 - 170) Ma and at the end (of 170 Ma to the present) of the Phanerozoic these quasi-periodic changes of the duration of polar intervals differed on average by 10% and were (5.5 Ma, 8 ma, and 15 Ma) and (5 Ma, 7 Ma, and 14 Ma), respectively. The details of the reconstruction of the geomagnetic reversals regime at the beginning and at the end of the Phanerozoic differ significantly. At the beginning of the Phanerozoic are gaps in the data on the behavior of the polarity of the geomagnetic field. Comparison of the results of reconstructing the geomagnetic polarity regime at the beginning and at the end of the Phanerozoic was shown that gaps of paleomagnetic data in the scale (Gradstein et al., 2008) have practically no effect on estimates of the duration of geomagnetic cycles.

**5. Common patterns of changes of paleointensity and duration of polar intervals**

As seen from Fig. 2 and Fig. 5, in the spectra of changes of paleointensity and the durations of geomagnetic polarity intervals there are identical quasi-periodicities with characteristic times of 15 Ma, 8 Ma, 5 Ma, and 3 Ma. In addition, the amplitude of paleointensity variations and duration of polar intervals increased if variations of geomagnetic activity were quasi-periodic, and their spectrum was discrete, for instance in the interval of (150 - 30) Ma. Changes in the paleointensity and duration of the polar intervals were chaotic in the interval of (170 - 150) Ma. At this time, both the average values of the amplitude of paleointensity variations and the average duration of the polar intervals decreased. The behavior of paleointensity and duration of polar intervals is characterized by two types of cyclical changes: 1) cyclical changes of the value of paleointensity and duration of polar intervals (with characteristic times of (3 - 15) Ma) and 2) cyclic changes in their spectral composition (according to our estimates, this process has characteristic times about 200 Ma).

It should be noted that some differences between the behavior of paleointensity and the duration of polar intervals are revealed. Clear connection between the amplitude of the paleointensity bursts and the duration of the polar intervals in which these bursts occurred there is no. Paleointensity bursts could occur during both relatively short and long intervals of geomagnetic polarity (Figs. 1 and 4).

Some differences are found in the dynamics of changes in the amplitude of the paleointensity bursts and the duration of the polar intervals. The maximum amplitude of paleointensity bursts was at the beginning of the Paleogene, and the longest polar intervals were in the middle Cretaceous. The decrease of the amplitude of paleointensity variations in the Paleogene began somewhat earlier than the decrease of the duration of the polar intervals near boundaries of 55 Ma and 30 Ma, respectively. From the data (Oxhneiser et al., 2013; Yamamoto et al., 2014 Yamazaki and Yamamoto, 2018), it follows that, before boundary of 30 Ma, variations with characteristic times exceeding the duration of individual polar intervals, are found in the behavior of paleointensity. After of 30 Ma, the characteristic times of paleointensity variations either exactly coincided with the duration of the polar intervals, or were shorter than the intervals of geomagnetic polarity. Such a conformity between paleointensity variations and the durations of polar intervals continued from 30 Ma to the present.

The detection of the same cyclicities in the behavior of paleointensity and geomagnetic polarity makes it possible to use on an equal basis the data on these parameters of the ancient geomagnetic field for conclusions about the duration of paleogeomagnetic cycles. Moreover, only paleointensity data make it possible to characterize the amplitude of changes of the activity of geomagnetic processes.

## 6. Discussion

We have analyzed three different arrays of paleomagnetic data: the geomagnetic polarity scale, and paleointensity data both by thermomagnetized and sedimentary rocks. The estimates of the duration of geomagnetic cycles (15 Ma, 8 Ma, 5 Ma, and 3 Ma) obtained from these arrays of paleomagnetic data practically coincide with each other. Such a coincidence of the estimates of the duration of geomagnetic cycles indicates that all these arrays of paleomagnetic data with an error of the order of 1 Ma adequately reflect the dynamics of changes of the ancient geomagnetic field. Thus, all modern paleomagnetic data make it possible to determine cyclical changes of the geomagnetic field with a resolution up to 3 Ma. In this case, the state of geomagnetic activity is directly determined only by the paleointensity behavior. Estimates of the spectrum of quasi-periodic changes of paleointensity are practically independent from the regions in which samples were taken for its determination. For instance, in the interval of (42 - 23) Ma paleointensity was studied using sediments that formed in different regions and were sampled with different details (Yamamoto et al., 2014; Kurazhkovskii et al., 2020). At the same time, the estimates of quasi-periodic changes of paleointensity made from these data coincide (Fig. 2a, 2b).

This study showed that the generation of the geomagnetic field is a complex process, during which both the intensity and the spectral composition of variations of the geomagnetic field strength were changed. Traditionally, it is assumed that the factors influencing the operation of a geodynamo are: 1) deep thermal processes, for example, the dynamics of heat transfer between the core and the mantle and 2) external processes during which the axial rotation rate of our planet changed (Braginsky, 1970). The duration of geomagnetic cycles, during which the values of paleointensity changed several times, was of (3 - 8) Ma. The relative changes in the intensity of thermal and rotational processes over such intervals of geological time were small in comparison with paleointensity changes. This indicates that between the factors that hypothetically affect on the generation of the geomagnetic field and the cyclical changes in its intensity there is no clear connection. Earlier, by Reshetnyak (2020) discussed the possibility of the influence of the redistribution of the energy of the geomagnetic field within the spectrum of its variations on the behavior of the main geomagnetic field. We have shown the nature of the relationship between the spectral composition of paleointensity variations and their amplitudes based on an analysis of paleomagnetic data.

Some difference in the estimates of spectral maxima at the beginning (5.5 Ma, 8 Ma, and 15 Ma) and at the end (5 Ma, 7 Ma, and 14 Ma) of the Phanerozoic can be associated with evolutionary changes of the medium in which the geomagnetic field was generated, and with gaps in paleomagnetic data. Since the detected changes in the duration of geomagnetic cycles are insignificant, there is a basis for the assumption that the physical properties of the medium in which the generation of the geomagnetic field took place during the Phanerozoic also changed insignificantly.

## 7. Conclusion

The identical cyclicities (quasi-periodicities) with characteristic times of 15 M, 8 Ma, 5 Ma, and 3 Ma were revealed in the spectra of changes of paleointensity and duration of polar intervals.

During the Phanerozoic (the last 550 Ma), the characteristic times of the geomagnetic cycles changed insignificantly. At the same time, the spectral density of quasi-periodic variations of paleointensity changed over the entire geological time. Quasi-periodic variations of paleointensity (geomagnetic activity) had a high spectral density in the interval of (150 - 40) Ma (in the Cretaceous - Early Paleogene). In this interval, the amplitude of changes of the paleointensity and the duration of the polar intervals increased. In the intervals (170 - 150) Ma and (of 30 Ma to the present) quasi-periodic changes of paleointensity practically did not detect out against the background of its noise variations. During the geomagnetic history, an alternation of time intervals in which paleointensity variations acquired either a quasi-periodic or noise character there was.

**Funding**

This study was carried out with the financial support of state task no. 0144-2014-00116.

**References**

Astaf'eva, N. M., 1996. Wavelet analysis: basic theory and some applications. Physics-Uspekhi. 39(11), 1085 - 1108. http://dx.doi.org/10.1070/PU1996v039n11ABEH000177.

Biggin, A.J., McCormack, A., Roberts, A., 2010. Paleointensity database updated and upgraded. EOS Trans. Am. geophys. Un. 91, 15. http://doi:10.1029/2010EO020003.

Bol'shakov, A.S., Solodovnikov, G.M., 1981. Geomagnetic field intensity in the last 400 million years. Dokl. Akad. Nauk SSSR. 260 (6), 1340 - 1343.

Braginsky, S.I., 1970. Fluctuation spectrum of the Earth's hydromagnetic dynamo. Geomagnetizm i aeronomiia. 10(3), 221 - 233. (in Russian).

Gradstein, F.M., Ogg, G.J., van Kranendonk, M., 2008. On the Geologic Time Scale 2008. News letters on stratigraphy. 43/1, 5 - 13. https://doi.org/10.1127/0078-0421/2008/0043-0005.

Guzhikov, A.Yu., Baraboshkin, E.Yu., Fomin, V.A., 2007. The Cretaceous magnetostratigraphic scale: state of the art, problems, and outlook, in: Pervushov, E.M. (Ed.), The Cretaceous System of Russia and FSU Countries: Problems of Stratigraphy and Paleogeography (in Russian). Saratov University Press, Saratov, pp. 69 - 86.

Heller, R., Merrill, R. T., McFadden, P. L., 2003. The two states of paleomagnetic field intensities for the past 320 million years. Phys. Earth Planet. Inter. 135, 211 - 223. https://doi.org/10.1016/S0031-9201(03)00002-5.

Hounslow, M.W., Domeier, M., Biggin, A. J., 2018. Subduction flux modulates the geomagnetic polarity reversal rate. Tectonophysics. 742–743, 34 - 49. https://doi.org/10.1016/j.tecto.2018.05.018.

Khramov, A.N. (Ed.), 1982. Paleomagnetology (in Russian), Nedra, Leningrad, 312 pp.

Kulakov, E.V., Sprain, C. J., Doubrovin, P.V., Smirnov, A. V., Paterson, G. A., Hawkins L., Fairchild, L., Piispa, E. J., Biggin, A. J., 2019. Analysis of an Updated Paleointensity Database ($Q_{PI}$-PINT) for 65 - 200 Ma: Implications for the long - term history of dipole moment through the Mesozoic. J. Geophys. Res. Solid Earth. 124. https://doi.org/10.1029/2018JB017287.

Kurazhkovskii, A.Yu., Kurazhkovskaya, N.A., Klain, B.I., Bragin, V.Yu., 2010. The Earth's magnetic field history for the past 400 Myr. Russian Geology and Geophysics. 51(4), 380 - 386. https://doi.org/10.1016/j.rgg.2010.03.005.

Kurazhkovskii, A.Yu., Kurazhkovskaya, N.A., Klain, B.I., 2011. Calibration of geomagnetic paleointensity data based on redeposition of sedimentary rocks. Phys. Earth Planet. Inter. 189, 109 - 116. https://doi.org/10.1016/j.pepi.2011.08.002.

Kurazhkovskii, A.Yu., Kurazhkovskaya, N.A., Klain, B.I., 2020. Paleointensity Bursts in the Geomagnetic Field History. Geomagnetism and Aeronomy. 60(6), 793 - 799. https://doi.org/10.1134/S0016793220050114.

Larson, R.L., Olson, P., 1991. Mantle plumes control magnetic reversal frequency. Earth Planet. Sci. Lett. 107, 437 - 447. https://doi.org/10.1016/0012-821X(91)90091-U.

MATLAB, Wavelet Toolbox Documentation:

https://www.mathworks.com/help/wavelet/index.html?s_tid= CRUX_lftnav. Accessed December 17, 2018.

Mazaud, A., Laj, C., Laurent de Seze, Verosub, K.L., 1983. 15 - Myr periodicity in the frequency of geomagnetic reversals since 100 Myr. Nature. 304, 328 - 330. https://doi.org/10.1038/304328a0.

Ohneiser, C., Acton, G., Channell, J.E.T., Wilson, G.S., Yamamoto, Y., Yamazaki, T., 2013. A middle Miocene relative paleointensity record from the Equatorial Pacific. Earth and Planet. Sci. Lett. 374, 227 - 238. http://dx.doi.org/10.1016/j.epsl.2013.04.038.

Pimenov, M.V., Yampol'skaya, O.B., 2008. Composite magnetostratigraphic column of the Middle–Upper Jurassic of the Russian Plate, in: Staroverov, V.N. (Ed), Essays on Regional Geology: to the 70th Anniversary of the Department of General Geology and Minerals Resources of the Faculty of Geology and 100th Anniversary of the N. G. Chernyshevsky Saratov State University (in Russian). Publishing Center "Science", Saratov, pp. 68 - 81.

Reshetnyak, M.Y., 2020. Evolution of the large-scale geomagnetic field over the last 12000 years. Geomagnetism and Aeronomy. 60, 121 - 132. https://doi.org/10.1134/S0016793220010119.

Serebrennikov, M.G., Pervozvansky, A.A., 1965. Revealing hidden periodicities (in Russian), Publishing House "Science", Fizmatgiz, Moscow, 244 pp.

Yamamoto, Y., Yamazaki, T., Acton, G.D., Richter, C., Guidry, E.P., Ohneiser C., 2014. Palaeomagnetic study of IODP Sites U1331 and U1332 in the equatorial Pacific – Extending relative geomagnetic palaeointensity observations through the Oligocene and into the Eocene. Geophys. J. Int., 196(2), 694 - 711. https://doi.org/10.1093/gji/ggt412.

Yamazaki, T., Yamamoto, Y., 2018. Relative paleointensity and inclination anomaly over the last 8 Myr obtained from the Integrated Ocean Drilling Program Site 1335 sediments in the eastern equatorial Pacific. J. Geophys. Res. Solid Earth. 123, 7305 - 7320. https://doi.org/10.1029/2018JB016209.

Zhamoida, A.I., Kovalevskii, O.P., Koren', T.N., Margulis, L.S., Predtechenskii, N.P., Rublev, A.G., Semikhvatov, M.A., Khramov, A.N., Shkatova, V.K., 2000. Additions to the Stratigraphic Code of Russia (in Russian). VSEGEI, St. Petersburg, 112 pp.